\documentclass[11pt]{article}

\usepackage[margin=1in]{geometry}
\usepackage{graphicx}
\usepackage{amsmath,amssymb}
\usepackage{authblk} 
\title{Fabry-Perot interferometer with quantum well mirror for controllable dispersion compensation}

\author[1]{Victor N. Mitryakhin}
\author[1]{Pavel Yu. Shapochkin}
\author[1]{Roman S. Nazarov}
\author[1]{Yury P. Efimov}
\author[1]{Sergey A. Eliseev}
\author[1]{Vyacheslav A. Lovcjus}
\author[1,*]{Yury V. Kapitonov}

\affil[1]{St. Petersburg State University, ul. Ulyanovskaya 1, Saint Petersburg, Peterhof, Russia}

\affil[*]{Corresponding author: yury.kapitonov@spbu.ru}

\begin{document}

\maketitle

\begin{abstract}
In this work, we investigate a possibility of controlling second-order dispersion in a monolithic Fabry-Perot interferometer based on epitaxial heterostructure with quantum well (QW) serving as a bottom mirror. Careful choice of heterostructure parameters and experimental conditions makes it possible to introduce negative dispersion in a very narrow spectral region of QW excitonic resonance while maintaining a constant reflection coefficient across this region. The feasibility of the concept is demonstrated for heterostructures with InGaAs/GaAs QWs at cryogenic temperatures. We also propose an active device design that can switch the dispersion compensation on and off by controlling the exciton ensemble’s environment.
\end{abstract}

Dispersion control plays an important role in optical systems where normal dispersion in the transparency region of their composition starts to significantly act on the performance of these devices. In this regard, common consequences of the latter include signal distortion and loss at long distances in optical communication systems \cite{Orfanidis2016} or pulse width broadening and coherence degradation of the output of short-pulsed lasers \cite{French1986, Sibbett1993}. In most cases, these effects can be suppressed via introduction of additional negative (anomalous) dispersion, determined by negative group delay dispersion (GDD), expressed as  $D_2(\omega) = \frac{\partial^2 \Phi}{\partial \omega^2}$, where $\Phi$ is the optical phase. In many applications, specific interferometer schemes, e.g. Fabry-Perot interferometer (FPI)-based~\cite{FabryPerot1899} reflectors, are used to achieve these goals. Such resonators are simple optical cavities, which are formed by two parallel mirrors. The complex reflection coefficient $r$ can thus be expressed as follows:

    \begin{equation}\label{fr_ref_main}
        r=\dfrac{r_{1}+r_{2}e^{2 i \varphi}}{1+r_{1}r_{2}e^{2 i \varphi}},
    \end{equation}

\noindent where $r_1$ and $r_2$ are complex reflection coefficients of mirrors and $\varphi$ is the phase build up after a round-trip in the cavity.

One scheme similar to the FPI is the Gires-Tournois interferometer~\cite{GTI1964} (GTI), which differs from the former by having an asymmetric design featuring a highly reflective second mirror ($|r_2|^2\approx1$) and thus primarily operating in reflection. Notably, in such a scheme the compensation of chromatic dispersion could be implemented in a lossless manner, since the reflectivity $K_R = |r|^2$ of the GTI is close to unity at any wavelength within the spectral stopband of the second mirror. The bandwidth of operation can be further extended by implementing more complex designs such as multi-cavity GTIs \cite{Szipocs2000}. In turn, the phase of reflected light $\Phi = \arctan \frac{\mathrm{Im}(r)}{\mathrm{Re}(r)}$ depends on the light frequency in a periodic manner. The GTI has a spectral region, where it may introduce fixed negative GDD without the possibility to control its value except for complex variable-gap GTI schemes that are subject to drift \cite{Trubetskov:13}.

Another common option for GDD control is a saturable absorber semiconductor mirror (SESAM), which is a GTI with a saturable absorber introduced into the gap between its mirrors. SESAM could be used to introduce different negative GDD values depending on the incident light intensity. Additionally, such active mirrors could be used for mode-locking and short pulse generation in lasers \cite{Keller1996, Isomaki2003, Isomaki2004, Moenster2007}.

In this work we propose new implementation of an active mirror for dispersion control -- an FPI-like monolithic structure with the quantum well (QW) acting as one of the mirrors. Design of the structure and the excitation geometry results in wavelength-independent reflectivity with the negative GDD introduced near the QW exciton resonance. Such FPI is similar to the GTI, but allows one to switch the GDD to zero by  bleaching the exciton resonance (for example, by additional control pulse). We theoretically describe the operation of the device, explore and demonstrate the feasibility of its design via the reflectivity measurements of the heterostructure with an InGaAs/GaAs QW.

\section{Principle of operation}

\begin{figure}
        \centering
        \includegraphics[width=0.85\linewidth]{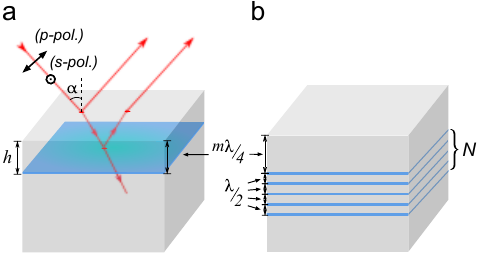} 
        \caption{\label{fig_structure} Single (a) and multiple (b) QW heterostructures under consideration. Red lines represent light beams paths. $\lambda$ -- wavelength in the material medium, $\alpha$ -- angle of incidence, $h$ -- cap layer thickness.}
    \end{figure}

Let us consider a heterostructure depicted in Fig. \ref{fig_structure}a and consisting of a substrate, a thin QW layer, and a cap layer with thickness~$h$. The refractive index of the semi-infinite substrate and the cap layer is $n_2$, and the air refractive index is $n_1 < n_2$. In this case one can regard the surface and the QW planes as two mirrors much like in the FPI design. The amplitude reflection coefficients of structure/air interface and of the QW layer denoted as $r_1$ (real number) and $r_2$ (complex number) correspondingly. Building up on Eq.~(\ref{fr_ref_main}), the reflectivity of the structure $K_R = |r|^2$ could be represented as follows:

    \begin{equation}
        \label{fr_ref_sep}
        K_R=
        |r_{1}|^2
        +
        \frac{
                \left( 1 - |r_1|^2 \right)
                \left(
                    |r_1|^2 + 2r_1 \mathrm{Re} \left( \frac{r_2}{|r_2|^2} e^{2 i \varphi} \right) + 1 
                \right) 
            }
            {
                |r_1|^2 + 2r_1 \mathrm{Re} \left( \frac{r_2}{|r_2|^2} e^{2 i \varphi} \right) + \frac{1}{|r_2|^2}
            }.
    \end{equation}

In the case of a thin single QW the reflection coefficient $r_2$ is expressed as such~\cite{Ivchenko2005}:

    \begin{equation}\label{fr_ref}
        r_2 = \frac{i\Gamma}{\Delta\omega-i(\Gamma + \gamma)},
    \end{equation}
    
\noindent where $\Gamma$ is the exciton resonance radiative width, $\gamma$ -- its non-radiative broadening, and $\Delta \omega = \omega_0 - \omega$ -- detuning of the exciton resonance frequency $\omega_0$ from the light frequency $\omega$.

The proposed FPI structure, as stated earlier, should have constant reflectivity across the whole spectral range. In this regard, such a situation will be referred to as compensation mode. 
Away from the exciton resonance, the non-resonant Fresnel reflectivity $|r_1|^2$ is the main contribution to the $K_R$. Closer to the excitonic resonance, its contribution to the reflectivity becomes notable and, thus, to maintain the spectral uniformity, it is necessary for the second term in Expression~(\ref{fr_ref_sep}) to be equal to zero across the whole spectral range. In this case, it can be achieved when the following condition is met (taking into account Eq.~(\ref{fr_ref})):

    \begin{equation}
    \label{cond_qw}
        r_{1}^2
        -
        2r_{1}\left(\frac{\Delta\omega}{\Gamma}\sin{2\varphi}+\left(1+
        \frac{\gamma}{\Gamma}\right)\cos{2\varphi}\right)+1=0.
    \end{equation}

This condition could be satisfied for any $\Delta \omega$ only if $\sin 2 \varphi = 0$. Thus, the necessary criterion for the constant reflectivity is $h = \frac{m \lambda}{4 \cos{\alpha} }$, where $m = 0, 1, 2, ...$. Note that this is valid only for sufficiently small $h$, when the phase $\varphi$ can be considered weakly dependent on the wavelength. Taking this criterion into account, Eq.~(\ref{cond_qw}) could be solved leading to the condition $r_1 = (-1)^m r_c$, where the compensation reflection coefficient is the following:  

    \begin{equation}
    \label{r_c}
        r_{c}= \dfrac{1}{F} - \sqrt{\left(\dfrac{1}{F}\right)^2 - 1}, 
    \end{equation}

\noindent while $F = \frac{\Gamma}{\Gamma + \gamma}$ is a quality factor and $r_c \ge 0$. For ideal QW ($\gamma = 0$): $F = 1$, while for realistic QW structures: $F < 1$.

For a given QW cap layer thickness (determined by $m$) and quality factor $F$ the compensation mode can always be found for certain light polarization and incident angle. Compensation conditions are summarized in Tab.~\ref{cond_table}, where $r_n = r_1(\alpha=0)$ is the reflection coefficient for structure/air interface at normal incidence and $\alpha_{Br} = \arctan \frac{n_2}{n_1}$ is Brewster's angle.

    \begin{table}
        \centering
        \caption{\bf Compensation conditions. \label{cond_table}}
        \begin{tabular}{ccccc}
        \hline 
        $m$ & Quality factor & Angle of incidence & Polarization \\
        \hline
        even & -- & $\alpha \geq \alpha_{Br}$ & $p$ \\
        odd  & $F < \dfrac{2r_{n}}{1+r_{n}^{2}}$ & $\alpha < \alpha_{Br}$ & $p$ \\
        odd  & $F = \dfrac{2r_{n}}{1+r_{n}^{2}}$ & $\alpha = 0$ & $p$, $s$ \\
        odd  & $F > \dfrac{2r_{n}}{1+r_{n}^{2}}$ & $\alpha > 0$ & $s$ \\
        \hline
        \end{tabular}
    \end{table}

\begin{figure*}
        \centering
        \includegraphics[width=0.85\linewidth]{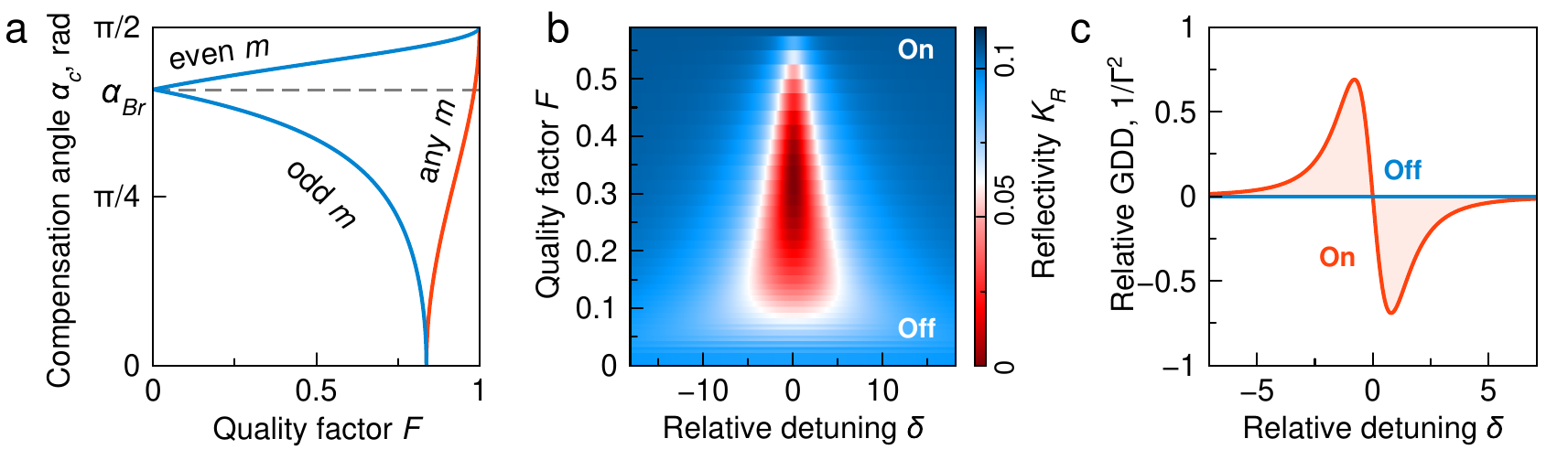}
        \caption{\label{fig_calc} 
            (a) Compensation mode polarization and incidence angles $\alpha_c$ for different quality factors $F$ for $n_1 = 1$ and $n_2 = 3.36$ (GaAs). Blue curve -- p-polarization, red -- s-polarization. Note that the compensation is always possible for any $m$ but at different polarizations and $\alpha_c$.
            (b) Reflectivity $K_R$ spectra for different quality factors $F$ demonstrating the same reflectivity for ''on'' state (compensation) and ''off'' state ($F=0$).  
            (c) GDT spectrum for ''on'' and ''off'' states shown in (b).
            Calculation parameters for (b) and (c): $n_1 = 1$, $n_2 = 3.36$ (GaAs), p-polarization, incident angle $\alpha = \alpha_c = 55.4^\circ$, $\Gamma = 86$~$\mu$eV, $\gamma$ is a variable.
        }
    \end{figure*}

Note that at $F = \dfrac{2r_{n}}{1+r_{n}^{2}}$ and odd $m$ the compensation takes place at normal incidence ($\alpha_c = 0$), and thus does not depend on the polarization of the incident light. The angle of incidence $\alpha_c,$ at which the compensation occurs, can be expressed as follows: 

    \begin{equation}
        \label{comp_angles}
            \alpha_c = \arccos{\sqrt{\dfrac{f-1}{f_{s, p}-1}}},
    \end{equation}

\noindent where $f = \dfrac{n^2_2}{n^2_1}$, $f_s=\left(\dfrac{1-r_1}{1+r_1}\right)^2$ and $f_p = \dfrac{f^2}{f_s}$ for s- and p-polarizations correspondingly. $\alpha_c(F)$ is shown in Fig.~\ref{fig_calc}a.

The compensation at normal incidence, which is particularly interesting for practical applications, imposes rather strong requirements on the QW quality. For example, in the case of GaAs/vacuum interface ($n_1 = 1$, $n_2 \approx 3.6$) the value of quality factor F for the normal compensation has to be equal to approximately 0.86, which is beyond the capabilities of current QW growth methods. This requirement can be relaxed by substituting the single QW design with a multiple QW (MQW) structure (Fig.~\ref{fig_structure}b). The MQW structure consists of $N$ QW layers, which are separated by $\lambda / 2$, in order to have the reflections from different QW interfaces interfering constructively\cite{Ivchenko2005}. If all the QWs have approximately the same $\gamma$ and $\Gamma$ values, all the earlier calculations remain the same. However, the single QW quality factor $F$ has to be replaced by effective quality factor $N \cdot F_0$, where $F_0$ is the quality factor of a single QW in the MQW structure. In this case the compensation at normal incidence still could be observed for a MQW structure composed of low-quality QWs. Notably, $F > 1$ is also achievable for MQW structures, although in this situation the compensation becomes impossible.


The reflection phase $\Phi$ is defined as $\tan{\Phi}=\dfrac{ \mathrm{Im}(r)}{\mathrm{Re}(r)}$. Expressing the reflection phase for the structure with $h = \frac{m \lambda}{4 \cos \alpha}$ and using the compensation condition $r_1 = (-1)^m r_c$ in the vicinity of the resonance frequency $\omega_0$ results in:

    \begin{equation}
        \Phi
        =
        \arctan{
        \left(
            \frac{
                \mathrm{Im}(r_2) \left( 1-|r_c|^2 \right)
            }{
                \mathrm{Re}(r_2) \left( 1 + |r_c|^2\right)
                +
                r_c \left( 1 + |r_2|^2\right)
            }
        \right)    
        }.
    \end{equation}

The substitution of QW parameters from (\ref{fr_ref}) and (\ref{r_c}) produces the following expression:

    \begin{equation}
            \label{comp_Phi}
            \Phi=\arctan
            \left(
                 \frac{ 2 \delta \xi }{ \delta^2-\xi^2 }
            \right),
    \end{equation}

where $\delta = \frac{\Delta\omega}{\Gamma}$ is the relative detuning and $\xi = \sqrt{(1/F)^2 - 1}$.

The group delay time (GDT) can be expressed in units of $\frac{1}{\Gamma}$ as a derivative of the reflection phase with respect to the relative detuning $\delta$:
    
    \begin{equation}
        \frac{\partial\Phi}{\partial\delta}
        =
        -
        \frac{2 \xi}
        {\delta^2 + \xi^2}.
    \end{equation}
    
As a result, the GDT is always negative. One more differentiation step returns the GDD in $\frac{1}{\Gamma^2}$ units:
    
    \begin{equation}
        \frac{\partial^2 \Phi}{\partial\delta^2}
        =
        \frac{4 \delta \xi}
        {\left( \delta^2 + \xi^2 \right)^2}.
    \end{equation}
    
\begin{figure}
    \centering
    \includegraphics[width=1.0\linewidth]{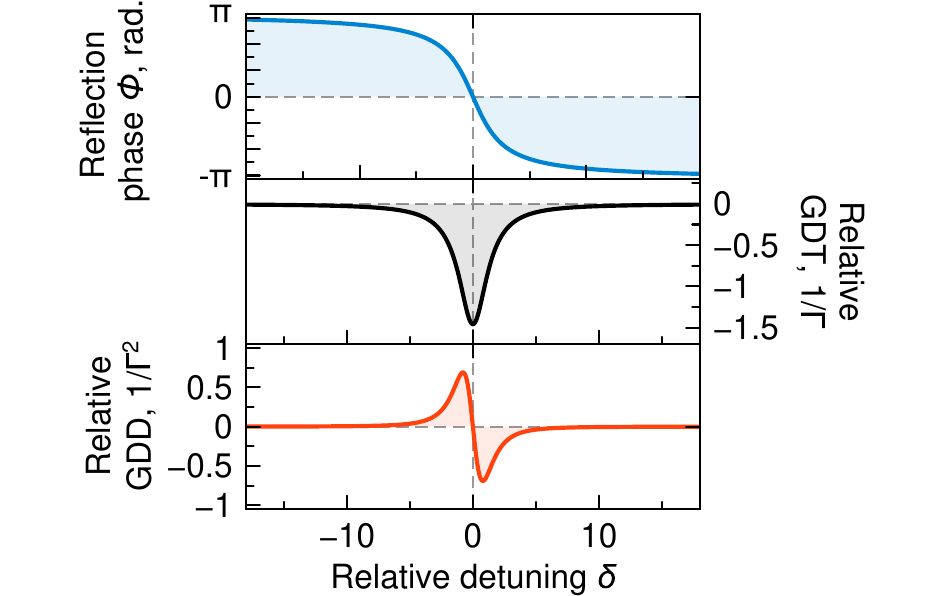} 
    \caption{\label{fig_abphi} Calculated reflection phase $\Phi$ (blue) and its derivatives: relative group delay time GDT (red) and relative group delay dispersion GDD (orange) in reflection compensation mode. Calculation parameters: $n_1 = 1, n_2 = 3.36$, p-polarization, incident angle $\alpha = \alpha_c = 55.4^\circ$, $F = 0.58$.}
\end{figure}

Fig.~\ref{fig_abphi} shows phase, GDT and GDD values for $F = 0.58$ (same as in the QW structure studied in the experimental section).
    
The advantage of the proposed QW-based mirror is the ability to switch on/off the dispersion compensation by an external stimulus. Fast bleaching of the QW resonance by a control laser beam can decrease the value of $F$ to almost zero. This could be done by rapid increase of $\gamma$ due to the excitons scattering on quasiparticles introduced by the control light pulse~\cite{Shapochkin2019, Shapochkin2020,  
Kurdyubov2021, Gribakin2021}. Fig.~\ref{fig_calc}b shows the reflectivity spectra for different $F$ values. The compensation in the ''on'' state takes place at $F = 0.58$ with the constant $K_R$ and non-zero GDD (Fig.~\ref{fig_calc}c, red curve). The mirror could be switched to the ''off'' state by the rapid decrease of $F$ to zero. In this case the reflectivity remains the same, but the GDD vanishes (Fig.~\ref{fig_calc}c, blue curve).

\section{Experiment}

To demonstrate the compensation mode, a sample with QW was grown using molecular beam epitaxy. The 2~nm thick In$_{0.02}$Ga$_{0.98}$As/GaAs QW has a GaAs cap layer of a 295~nm target thickness and was grown on a (100) GaAs substrate overgrown by a GaAs barrier. The thickness of the cap layer is roughly equal to $\frac{5 \lambda}{2}$ ($m = 5$). 

The sample was grown without substrate rotation, which resulted in a gradient in the cap layer thickness across the sample on a millimeter scale. The optical probe spot was below 100~$\mu$m in diameter. Therefore, it is possible to locate a position on the sample where the cap layer thickness $h$ satisfies the compensation condition. The optical thickness $h$ can also be tuned by choosing a specific angle of incidence $\alpha = \alpha_c$. In our experiment we have fixed $\alpha < \alpha_{Br}$ and investigated the case of $p$-polarization with the compensation condition presented in the second row of Tab.~\ref{cond_table}). 

The sample was kept in a closed-loop helium cryostat during the optical studies. We probed the reflectivity of the structure using a white halogen lamp focused to a spot of 100~$\mu$m in diameter and a spectrometer with a CCD detector. The thickness of the cap layer was rather uniform across the spot area due to the weak gradient of the cap layer thickness. 

We studied the QW heavy-hole exciton resonance observed at $1.5125$~eV at $T=$~10~K. The shape of the observed QW exciton resonance in the reflection spectra depends on the phase of the reflection from the QW plane and, thus, the thickness of the cap layer. For a thickness equal to a multiple of $\frac{\lambda}{4}$ the shape is symmetric. Once the compensation condition (\ref{r_c}) is satisfied, the exciton peak area over the background reflectivity level approaches zero. 

To fine-tune the operation mode of the proposed device, one has to consider manipulating exciton ensemble properties (the quality factor $F$) as required by expression (\ref{r_c}). Since $\Gamma$ of the exciton resonance is fixed for the QW composition and the thickness \cite{Poltavtsev2014}, the quality factor can be varied by tuning $\gamma$. Non-radiative broadening $\gamma$ is caused by non-radiative relaxation pathways including thermally mediated processes (exciton-phonon interaction) and all other processes of the irreversible phase relaxation, such as exciton scattering on charge carriers, excitons and other elementary excitations. In this work, we introduce $\gamma$ change in the experiment via the increase of the homogeneous broadening by two distinct methods: the additional illumination of the sample above the QW barrier (in our case, by a 656~nm CW-laser) and the heating of the sample. 

\begin{figure}
    \centering
    \includegraphics[width=1\linewidth]{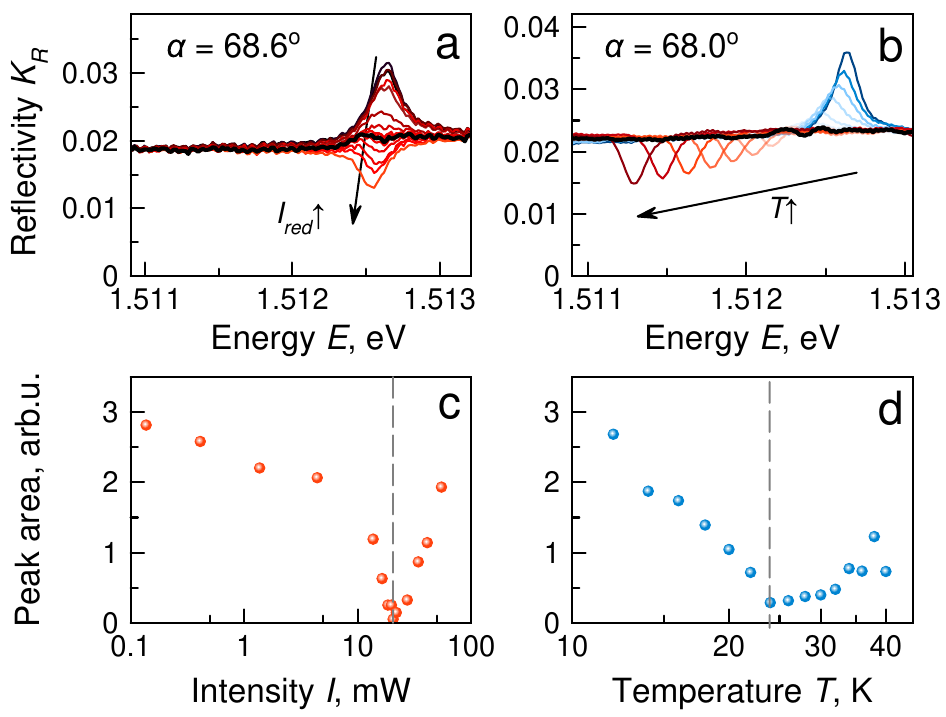} 
    \caption{
    \label{fig_exp}
        Experimental reflectivity spectra $K_R(E)$ (a,b) and dependence of the area of the heavy-hole exciton peak (c,d) on additional illumination at $T=10$~K (a,c) and on sample temperature (b,d). Black curves in (a,b) denote the reflectivity when the compensation condition is met at dashed lines position in (c,d).
    }
\end{figure}

Fig.~\ref{fig_exp}a shows the reflectivity of the sample at $T=$~10~K for different intensities of the additional 656~nm illumination up to 100~mW. The illumination introduces additional homogeneous broadening due to the exciton-exciton and exciton-carriers scattering, and thus decreases $F$. In the absence of illumination, constructive interference of signals reflected from the QW and the sample surface is observed. With an increase in the illumination intensity $I$, the amplitude of the exciton peak decreases, and at a certain intensity, the compensation is observed -- the reflectivity $K_R$ is constant (black curve). With a further increase of $I$ the interference becomes destructive and the exciton peak inverts. Fig.~\ref{fig_exp}c shows the peak area dependence on the intensity of the additional irradiation~$I$. At $I=20$~mW the compensation is observed as indicated by a minimum in the peak area.

A similar experiment was carried out while the sample temperature was swept from 10 to 40~K (Fig.~\ref{fig_exp}b). An increase in temperature~$T$ led to an increase in the thermal broadening, and also a red shift of the exciton resonance due to the energy bandgap of the material changing. At $T=24$~K the minimum of the interference peak was reached (see Fig.~\ref{fig_exp}d) and the compensation condition was met.

The following parameters were obtained from the fitting: compensation is observed at $F = 0.58$; the maximum GDT is -10~ps; maximum absolute value of GDD is $3.6 \cdot 10^6$~fs$^2$


\section{Conclusion}

In this work, we have proposed the method of controlled dispersion compensation using a Fabry-Perot interferometer with a QW mirror. We have theoretically and experimentally demonstrated that it is possible to find a specific configuration in which there is no reflection feature associated with exciton resonance and the reflectivity remains constant. The presence of a non-zero GDD near the exciton resonance was predicted. This GDD could be actively switched on and off by an external stimulus without introducing any change to the reflectivity spectrum. The proposed MQW design will allow us to achieve dispersion compensation at normal incidence and obtain larger GDD values. The reflectivity close to unity could be achieved by an inverted design with a top QW mirror and a bottom distributed Bragg reflector, similar to the design in Ref.~\cite{Isomaki2003}. In the future, the proposed devices could be designed to operate at room temperature by switching to other exciton materials with higher exciton binding energy, such as GaN 
or 2D van-der-Waals materials. 

\section{Funding}

This research has been supported by Russian Science Foundation and St.Petersburg Science Foundation grant No. 25-12-20007 (https://rscf.ru/project/25-12-20007/).

\section{Acknowledgments}

The study was carried out using the equipment of Saint Petersburg State University resource center ''Nanophotonics''.

\section{Disclosures}

The authors declare no conflicts of interest.

\bibliography{literature}

This is the author-accepted version of a paper published in Optics Letters. 
The final version is available at DOI: 10.1364/OL.572092.

\end{document}